%% file: ms.tex
\documentclass[12pt,twoside,english]{article}
\usepackage{ifthen}                
\newboolean{boldeqnitalic}
\setboolean{boldeqnitalic}{true}

\usepackage{caption}
\usepackage[intlimits] {amsmath}   
\usepackage{defjs2}                
\usepackage{epsfig}
\usepackage{psfig}
\usepackage{graphicx}
\usepackage{psfrag}
\usepackage[usenames]{color}
\usepackage{colortbl}
\usepackage{ifthen}
\usepackage{fancyhdr}
\usepackage{rotating}
\usepackage{multirow}
\usepackage{booktabs}              
\usepackage{amssymb}
\usepackage{amsbsy}
\usepackage{bm}                    
\usepackage{babel}
\usepackage{theorem}
\usepackage{upgreek}  
\usepackage{pifont}   
\usepackage{mathrsfs}
\usepackage{datetime}
\usepackage{tikz}
\usetikzlibrary{matrix}
\usepackage[all]{xy}
\usepackage{rotating}
\usepackage{subfigure}

\definecolor{dunkelgrau}{rgb}{0.8,0.8,0.8}
\definecolor{hellgrau}{rgb}{0.90,0.90,0.90} 
\usepackage[authoryear]{natbib}   
\fancypagestyle{plain}{%
  \fancyhead{}%
  \fancyfoot[c]{\sffamily\thepage}%
}
\makeatletter                           
\def\cleardoublepage{\clearpage\if@twoside \ifodd\c@page\else
  \hbox{}
  \vspace*{\fill}
  \thispagestyle{empty}
  \newpage
  \if@twocolumn\hbox{}\newpage\fi\fi\fi}
\makeatother
\begin{document}
\unitlength1.0cm
\frenchspacing

\include{antrag_text}
\bibliographystyle{plainnat}
\end{document}

%% file: antrag_text.tex
\thispagestyle{empty}
 
\input{Frontpage-1}

 
\section{Introduction}
\label{sec:intro}
 
Diamond thin films find an every more increasing interest for their outstanding mechanical properties, most notably for their high hardness, low friction coefficient and high wear resistance. A critical issue for their industrial application is their poor adhesion on many substrates due to high interface stresses induced by different thermal expansion coefficients of the diamond film and the substrate. An additional problem is the catalytic effect in the context of iron-, cobalt- and nickel-based materials which results in soot and graphite formation. As an effective solution to these problems a nanocrystalline diamond/$\beta$-SiC composite film is used as a transition layer that serves the purpose of an adhesion layer and/or a barrier layer to prevent the catalytic effect of the substrate elements. Diamond/$\beta$-SiC composite films fabricated by chemical vapor deposition (CVD) techniques open the door to create superior materials with combined advantages of both diamond and $\beta$-SiC for a wide range of applications, such as tribological and biological coatings, sensors \cite{Wang-etal2014}, \cite{Zhuang-etal-2010}, windows for harsh environment, electronic devices and many more. 

Since diamond/$\beta$-SiC composites are part of engineering structures in a multitude of applications, their mechanical performance is of cardinal importance for the functionality and integrity of structures at various length scales. Recent progress made in the synthesis, characterization and applications of the diamond/$\beta$-SiC composite films are reviewed in \cite{JiangZhuang2015}, an even broader view on current trends in diamond research and technology is given in \cite{Yang2015}. 
  
In view of the ever more increasing sophistication of fabrication and the rapidly growing field of applications there is a strong need for computational modeling diamond-based composites. Remarkably, this need is in stark contrast to the achievements of computational materials science for this type of composite material so far. One reason is the heterogeneity of the material; while for two-phase diamond/$\beta$-SiC composites the elastic material parameters of single phases are available, very little is known about the effective elasticity properties of the composites. 

The present work aims to bridge the existing gap by a thorough analysis of the effective elasticity properties through computational homogenization of a 3D representative volume element (RVE) which must be reconstructed from two sectional micrographs. Hence, the parameter identification has to be carried out coping with a lack of 3D data, as only planar microscopic images are available. 

\subsection{Fabrication and database}
\label{sec:AvailableData}

SiC exists in numerous crystalline forms (polytypes), $\beta$-SiC has cubic lattice which is the preferred type for composites with (cubic) diamond. The $\beta$ type is stable only at temperatures below 1700$^{\circ}$C, in contrast to $\alpha$-SiC which is formed at higher temperatures.

In the CVD fabrication of microcrystalline diamond/$\beta$-SiC composite thin films, single-crystalline p-type Si(100) wafers were used as substrates. A key requirement for the high crystallinity of diamond and $\beta$-SiC phases in the simultaneous co-deposition process is the utilization of high atomic hydrogen concentration. The high atomic hydrogen concentration, generated by a high microwave power density, hinders secondary heterogeneous nucleation and therefore enhances the selectivity of CH$_3$ and SiH$_3$ deposition onto diamond and $\beta$-SiC crystals. The high-quality diamond and $\beta$-SiC phase formation was examined and proved by scanning electron microscopy (SEM), high-resolution transmission electron microscopy (HRTEM), X-ray diffraction (XRD) and Raman spectroscopy, for more details of the fabrication process we refer to \cite{ZhuangZhangStaedlerJiang2011}.   

In this work the effective elastic properties of the composite material are computed based on a representation of the 3D microstructure and under consideration of the single phase properties (see Tab.\ref{tab:MatParam}) which each exhibit isotropic, linear elastic behavior. 3D microstructure information can be obtained with suitable tomography methods. For our composite material a voxel resolution of 5 nm  is required, which can be achieved with FIB-SEM tomography \cite{Holzer-etal-2004}, \cite{Holzer-Cantoni-2012}. Unfortunately, the acquisition of representative 3D-volumes is time consuming and costly. Materials optimization studies typically involve many different microstructures (e.g. by varying the CVD fabrication parameters). For such studies more efficient approaches for 3D characterization are required.

\begin{Table}[htbp]
\centering
\begin{tabular}{ccccc}
 \hline
 & Phase       &            & $E$ {[}GPa{]} & $\nu$   \\
 \hline
 & Diamond     &            & 775       & 0.20  \\
 & $\beta$-SiC &            & 250       & 0.17 \\
 \hline
\end{tabular}
\caption{Material parameters of isotropic, linear elasticity, Young's modulus $E$ and Poisson's ratio $\nu$ for each phase of the  diamond/$\beta$-SiC composite.}
\label{tab:MatParam}
\end{Table}

\cite{TurnerKalidindi2016} introduced statistical methods that enable 3D reconstruction of composite microstructures based on a set of 2D-images. The latter represent sections through the microstructure in different orientations. This approach thus characterizes the internal 3D microstructure without performing expensive tomography. In the present study we intend to apply this statistical 3D reconstruction approach as a basis for computation of elastic properties. Thereby, the number and size of 2D images is reduced to a minimum. The thus presented method can be used in the future for computational materials optimization with minimal efforts for experimental and imaging investigations, in particular for materials with transverse isotropy, which is often the case when using CVD.

The available database for estimating the 3D elasticity tensor in this work is restricted to two planar images of the two phase material; one for the surface plane, the second one for a cross section as displayed in Fig. \ref{fig:Plane-pics-Surf-and-CrossSect}. Notice that the bright phase is diamond, the dark phase SiC. 

\begin{Figure}[htbp]
	\centering   
	\includegraphics[width=0.99\linewidth]{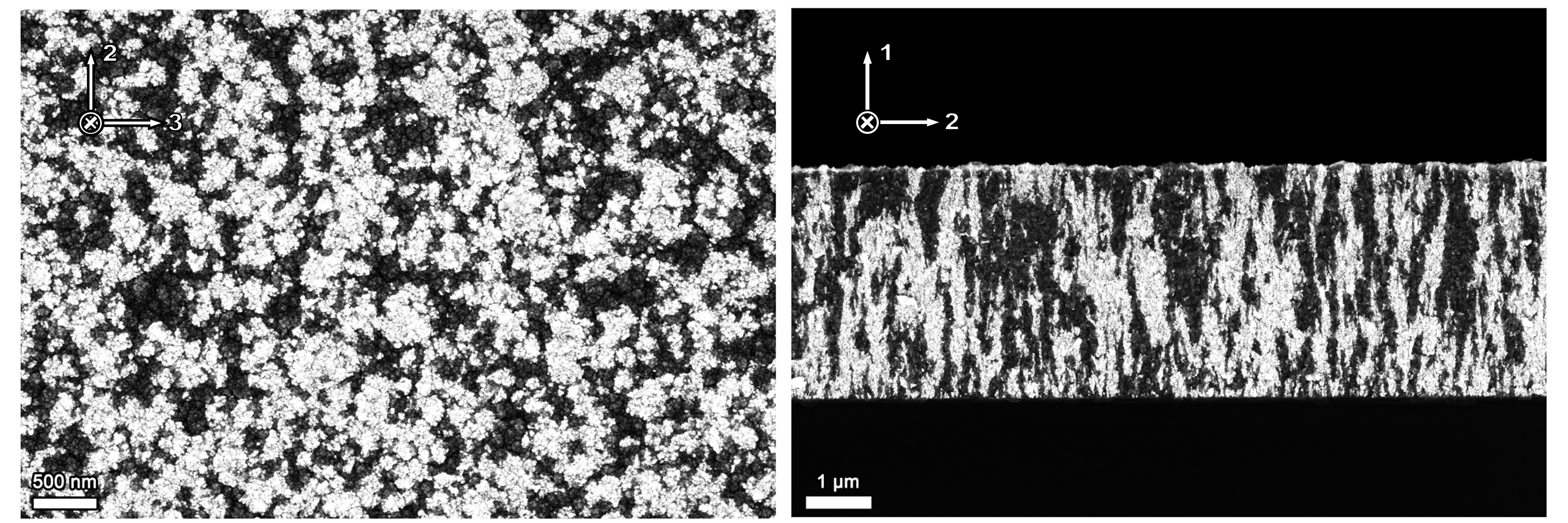}
	\caption{Diamond/$\beta$-SiC composite film: SEM-micrographs of (left) the surface plane and of (right) a cross sectional plane obtained {\color{black}by cutting, grinding and polishing}. Pixel resolution is (left) 1024 $\times$ 690 and (right) 1024 $\times$ 690.}
	\label{fig:Plane-pics-Surf-and-CrossSect}
\end{Figure}
  
The minimal number of planar cuts through the material for the 3D microstructure reconstruction method introduced in \cite{TurnerKalidindi2016} is three. Since only two cross-sectional cuts are available, the missing information on microstructure topology in the direction perpendicular to the existing planar data must be obtained by additional reasonable assumptions.   
 
\subsection{Modus operandi}
 
The following modus operandi for the reconstruction of an RVE and for the computation of the homogenized 3D elasticity tensor is pursued:
\begin{enumerate}
 \item Computation of the (i) phase fraction ratio and the (ii) homogenized 2D elasticity tensors, each for the surface plane and the cross sectional plane as displayed in Fig.\ref{fig:Plane-pics-Surf-and-CrossSect}. 
  \begin{enumerate}
    \item Consideration of full planar windows to cover the maximum of available microstructure information. 
    \item Consistency test, whether the phase fraction ratio for the surface plane equals the phase fraction ratio of the cross sectional plane.
    \item Test of the hypothesis that in the surface plane the type of 2D elasticity law is close to isotropy, and in the cross sectional plane close to orthotropy. Quantification of the deviation between the real, homogenized elasticity coefficients from isotropy and orthotropy, respectively.  
  \end{enumerate}
 \item Identification of 2D representative area elements (RAEs) of reduced size each for the surface plane and the cross sectional plane. The criteria for the RAE selection is statistical representativeness, i.e. matching (i) the phase fraction ratio as well as (ii) the elasticity properties of those of the full planar windows.
       Moreover, an exemplar for the missing third plane perpendicular to the slices in Fig. \ref{fig:Plane-pics-Surf-and-CrossSect} must be obtained to feed the reconstruction algorithm.     
 \item Reconstruction of a 3D RVE from RAEs by the application of an optimization software, \cite{TurnerKalidindi2016}. 
 \item Computation of the homogenized 3D elasticity matrix; identification of the type of anisotropic linear elasticity law showing the best fit to the homogenized elasticities, and determination of the corresponding material constants. 
\end{enumerate}

Arguments for and validation of the hypotheses of elastic isotropy in the surface plane and elastic orthotropy in the cross-sectional plane (point 1.(c) in the above list) will be given in the following.

\section{Results}  

\subsection{Planar pre-analyses}
In the first step the phase fractions are determined and the homogenized elasticity matrices for both the (23) surface plane as well as the (12) cross sectional plane (compare Fig. \ref{fig:Plane-pics-Surf-and-CrossSect-Window-Si}) are computed by numerical homogenization. For this purpose, the color code of the SEM image pixels is converted into binaries (black, white), where a threshold value for the color code distinguishes between the phases. 

In the planar pre-analyses we consider the full size of the SEM pictures shown in Fig.\ref{fig:Plane-pics-Surf-and-CrossSect} and thus cover the maximum of available information. For the 2D homogenization, plane stress conditions are assumed to hold.  
Periodic boundary conditions are imposed on the planar samples thus modeling an infinite extension of the exemplars in the plane. For linear elasticity (Hooke's law) stress and strain are related by a constant elasticity tensor $\mathbf{\mathbb C}$ as displayed in \eqref{Hookes law} using Voigt notation. Inversion of the symmetric elasticity matrix $\mathbf{\mathbb C}$ results in the compliance matrix $\mathbf{\mathbb S}$ with $\mathbb S = {\mathbb C}^{-1}$.

\begin{equation}
	\left[\begin{array}{c}
	\sigma_{11} \\ \sigma_{22} \\ \sigma_{33} \\ \sigma_{23} \\ \sigma_{13} \\ \sigma_{12}
	\end{array}\right] = \displaystyle \underbrace{\left(\begin{array}{c c c c c c}  \mathbb C_{11} &  \mathbb C_{12} & \mathbb C_{13} & \mathbb C_{14} & \mathbb C_{15} & \mathbb C_{16} \\  & \mathbb C_{22} & \mathbb C_{23} & \mathbb C_{24} & \mathbb C_{25} & \mathbb C_{26} \\  &  & \mathbb C_{33} & \mathbb C_{34} & \mathbb C_{35} & \mathbb C_{36} \\ & & & \mathbb C_{44} & \mathbb C_{45} & \mathbb C_{46} \\ & & & & \mathbb C_{55} & \mathbb C_{56} \\ \scriptstyle \text{sym.} & & & & & \mathbb C_{66}
	\end{array}\right)}_{\substack{\text{Elastic stiffness matrix } \mathbf{\mathbb C}}} \left[\begin{array}{c}\varepsilon_{11} \\ \varepsilon_{22} \\ \varepsilon_{33} \\ 2\varepsilon_{23} \\ 2\varepsilon_{13} \\ 2\varepsilon_{12}
	\end{array}\right]
\label{Hookes law}
\end{equation}
  
\begin{Table}[htbp]
	\centering
	\renewcommand{\arraystretch}{1.2}
	\begin{tabular}{c c c c c c c}
		\hline 
	    2D	& \multicolumn{6}{c}{Elasticity matrix coefficients}       \\
		\hline 
	   surf & $\mathbb C_{22}$& $\mathbb C_{33}$ & $\mathbb C_{44}$  & $\mathbb C_{23}$  & $\mathbb C_{24}$   & $\mathbb C_{34}$  \\
	        & $410.41$        & $408.69$         & $163.81$          & $78.13$           &         $ -0.24$   & $-0.76$           \\
	 	\hline
	  cross & $\mathbb C_{11}$  & $\mathbb C_{22}$ & $\mathbb C_{66}$  & $\mathbb C_{12}$  & $\mathbb C_{16}$  & $\mathbb C_{26}$     \\
 	        & $430.82$          & $384.04$ & $158.37$          & $73.97$           & $-0.49$           & $-0.57$     \\
		\hline 
	\end{tabular} 
	\caption{Coefficients of effective planar elasticity matrices (in GPa) for the (23) surface plane and for the (12) cross section covering the full sample sizes of the images in Fig.~\ref{fig:Plane-pics-Surf-and-CrossSect}.
	\label{tab:Homogenized-Elasticity-2d}}
\end{Table}

The resultant phase fraction ratios of diamond to SiC of 44:56 for the surface plane, and 42:58 for the cross sectional plane are in reasonable agreement with each other. The coefficients of the homogenized elasticity matrices --each from planar homogenization analyses-- are displayed in Tab.~\ref{tab:Homogenized-Elasticity-2d}. 
 
\subsubsection{Test for isotropy in the (23) surface plane.}
Since the surface plane in Fig.~\ref{fig:Plane-pics-Surf-and-CrossSect} does hardly exhibit any visible material direction, isotropy is a promising candidate for the type of elasticity in that plane. Moreover, the assumption is corroborated by the almost quantitative agreement of $\mathbb C_{22}$ and $\mathbb C_{33}$, which is an additional indicator for elastic isotropy, cf. Tab.~\ref{tab:Homogenized-Elasticity-2d}.

Isotropic elasticity is characterized by two independent material parameters, the Lam\'{e} constants $\lambda$ and $\mu$ or $E$ and $\nu$. Assuming the special case of plane stress and regarding only deformations in the corresponding (23) surface plane, the 2D elasticity law for isotropy follows the form of \eqref{2D isotropy elasticity law}, where only the non-vanishing coefficients of stress and strain vectors and the compliance matrix are given in full black color, whereas the vanishing terms are set in grey color.   

\begin{equation}
\left[\begin{array}{c}
\color{gray}{\varepsilon_{11}} \\ \varepsilon_{22} \\ \varepsilon_{33} \\ 2\varepsilon_{23} \\ \color{gray}{2\varepsilon_{13}} \\ \color{gray}{2\varepsilon_{12}}
\end{array}\right]=\displaystyle \underbrace{\left(\begin{array}{c c c c c c} \color{gray}{\mathbb S_{11}} & \color{gray}{\mathbb S_{12}} & \color{gray}{\mathbb S_{13}} & \color{gray}{\mathbb S_{14}} & \color{gray}{\mathbb S_{15}} & \color{gray}{\mathbb S_{16}} \\ & \frac{1}{E} &  -\frac{\nu}{E} & \color{gray}{\mathbb S_{24}} & \color{gray}{\mathbb S_{25}} & \color{gray}{\mathbb S_{26}} \\  & & \textstyle \phantom{-} \frac{1}{E} & \color{gray}{\mathbb S_{34}} & \color{gray}{\mathbb S_{35}} & \color{gray}{\mathbb S_{36}} \\  & & & \frac{2(1+\nu)}{E} & \color{gray}{\mathbb S_{45}} & \color{gray}{\mathbb S_{46}} \\ & & & & \color{gray}{\mathbb S_{55}} & \color{gray}{\mathbb S_{56}} \\ \scriptstyle \text{sym.} & & & & & \color{gray}{\mathbb S_{66}}
\end{array}\right)}_{\substack{\text{Elastic compliance matrix } \mathbf{\mathbb S} \\ \text{for isotropy}}} \left[\begin{array}{c}
\color{gray}{\sigma_{11}} \\ \sigma_{22} \\ \sigma_{33} \\ \sigma_{23} \\ \color{gray}{\sigma_{13}} \\ \color{gray}{\sigma_{12}}
\end{array}\right]
\label{2D isotropy elasticity law}
\end{equation} 

The parameters $E$ and $\nu$ are identified by the homogenized coefficients $\mathbb S_{22}$ and $\mathbb S_{23}$. The remaining  coefficients are used to validate the hypothesis of isotropy. Equation \eqref{2D isotropy elasticity law} implies 
\begin{center}
(i) $\mathbb S_{22}=\mathbb S_{33}$, (ii) $\mathbb S_{44}=1/G$, (iii) $\mathbb S_{24}=0$, (iv) $\mathbb S_{34}=0$\, , 
\end{center} 
\begin{center}
or, in terms of the stiffness matrix, 
\end{center} 
\begin{center} 
(i) $\mathbb C_{22}=\mathbb C_{33}$, (ii) $\mathbb C_{44}=G$, (iii) $\mathbb C_{24}=0$, (iv) $\mathbb C_{34}=0$.
\end{center}
Notice that for isotropy the shear modulus is not an independent material parameter; instead it holds $G=E/(2(1+\nu))$. 
 
These four conditions (i)-(iv) are used to check the homogenized elasticity tensor for isotropy. The deviation from isotropy is quantified by the percental deviation of the computed coefficients from the isotropic ones in \eqref{2D isotropy elasticity law} based on the identified parameters $E$ and $\nu$. For conditions (iii) and (iv) the deviation from isotropy is measured in terms of the magnitude of $\mathbb C_{24}$ and $\mathbb C_{34}$ (they both vanish for perfect isotropy) in comparison to the leading coefficient $\mathbb C_{22}$, see Tab.~\ref{tab:Deviation Isotropy}.
 
\begin{Table}[htbp]
	\centering
	\renewcommand{\arraystretch}{1.2}
	\begin{tabular}{ >{\centering\arraybackslash} m{0.01mm} >{\centering\arraybackslash} m{1cm} |  >{\centering\arraybackslash} m{1.5cm}  >{\centering\arraybackslash} m{1.5cm} |  >{\centering\arraybackslash} m{2cm}  >{\centering\arraybackslash} m{2cm}  >{\centering\arraybackslash} m{1.5cm} >{\centering\arraybackslash} m{1.5cm}  }
		\hline 
		 & & \multicolumn{2}{c}{Identification} & \multicolumn{4}{c}{Test for isotropy} \\
		 \hline 
	    \rule{0pt}{25pt} 
	      & 2D & $E$ & $\nu$ & $\displaystyle \frac{\mathbb C_{22}-\mathbb C_{33}}{\mathbb C_{22}}$ & $\displaystyle \frac{\vert \mathbb C_{44}-G \vert}{G}$ & $\displaystyle \frac{\vert \mathbb C_{24} \vert}{\mathbb C_{22}}$ & $\displaystyle \frac{\vert \mathbb C_{34} \vert}{\mathbb C_{22}}$    \\
		\hline 
	      & surf	& 395.47 & 0.191 & 0.42 $\%$  & 1.32 $\%$ & 0.06 $\%$  & 0.19 $\%$   \\
	 	\hline
	\end{tabular} 
	\caption{Identified elasticity constants (in GPa) of the entire surface plane assuming isotropy and deviation [$\%$] of the remaining coefficients from isotropy.} 	
	\label{tab:Deviation Isotropy}
\end{Table}

As a result, a comparison of coefficients underpins that the elasticity law in the (23) surface plane is very close to perfect isotropy thus justifying the hypothesis; the maximal deviation of a coefficient from the perfect isotropic one is less than 1.4\%, $\mathbb C_{24}$ and $\mathbb C_{34}$ virtually vanish. 
  
\subsubsection{Test for orthotropy in the (12) cross sectional plane.}
In view of the fabrication process by CVD and the resultant phase alternations of the diamond and SiC phase forming roughly a laminate in the (12) cross sectional plane, elastic orthotropy in that plane is justified as a working hypothesis.

For that case, \eqref{2D transverse isotropy law} displays the law of linear elasticity in terms of the compliance matrix $\mathbb{S}$. Four independent material parameters are enough to describe orthotropy in the plane, $E_1$, $E_2$, $\nu_{12}$ and $G_{12}$. The number of independent Poisson's contractions is reduced to one, here $\nu_{12}$, where the $1$-direction aligns with the growth direction of the phases, see Fig.~\ref{fig:Plane-pics-Surf-and-CrossSect-Window-Si}.  
 
\begin{equation} 
\begin{pmatrix}
\varepsilon_{11} \\ \varepsilon_{22} \\ \color{gray}{\varepsilon_{33}} \\ \color{gray}{2 \varepsilon_{23}} \\ \color{gray}{2\varepsilon_{13}} \\ 2\varepsilon_{12}
\end{pmatrix} = \underbrace{\left(\begin{array}{c c c c c c}
\frac{1}{E_1} & -\frac{\nu_{12}}{E_1} & \color{gray}{\mathbb S_{13}} & \color{gray}{\mathbb S_{14}} & \color{gray}{\mathbb S_{15}} & \color{gray}{\mathbb S_{16}} \\  & \phantom{-} \frac{1}{E_2} & \color{gray}{\mathbb S_{23}} & \color{gray}{\mathbb S_{24}} & \color{gray}{\mathbb S_{25}} & \color{gray}{\mathbb S_{26}} \\ & & \color{gray}{\mathbb S_{33}} & \color{gray}{\mathbb S_{34}} & \color{gray}{\mathbb S_{35}} & \color{gray}{\mathbb S_{36}} \\ & & & \color{gray}{\mathbb S_{44}} & \color{gray}{\mathbb S_{45}} & \color{gray}{\mathbb S_{46}} \\ & & & & \color{gray}{\mathbb S_{55}} & \color{gray}{\mathbb S_{56}} \\ \scriptstyle \text{sym.} & & & & & \frac{1}{G_{12}}
\end{array}\right)}_{\substack{\text{Elastic compliance matrix } \mathbf{\mathbb S} \\ \text{for orthotropy}}} \begin{pmatrix}
\sigma_{11} \\ \sigma_{22} \\ \color{gray}{\sigma_{33}} \\ \color{gray}{\sigma_{23}} \\ \color{gray}{\sigma_{13}} \\ \sigma_{12}
\end{pmatrix}
\label{2D transverse isotropy law}
\end{equation}

Notice that there are four conditions for a comparison of coefficients referring to the set $\{\mathbb S_{11}$, $\mathbb S_{12}$, $\mathbb S_{22}$, $\mathbb S_{66}\}$) to identify the four independent material parameters. The conditions 
\begin{center} 
 (a) $\mathbb S_{16}=0$, (b) $\mathbb S_{26}=0$ \qquad \mbox{or}  \qquad  (a) $\mathbb C_{16}=0$, (b) $\mathbb C_{26}=0$  
\end{center}
are left to check the validity of 2D orthotropy. Table \ref{tab:Deviation Orthotropy} displays the results of identification and validation; the almost vanishing magnitude of $\mathbb C_{16}$ and $\mathbb C_{26}$ in comparison to the leading coefficient $\mathbb C_{11}$ (less than 0.2\%) underpins the excellent agreement with elastic orthotropy. 

\begin{Table}[htbp]
	\centering
	\renewcommand{\arraystretch}{1.2}
	\begin{tabular}{ >{\centering\arraybackslash} m{0.01mm} >{\centering\arraybackslash} m{1.3cm} | >{\centering\arraybackslash} m{1.2cm} >{\centering\arraybackslash} m{1.2cm} >{\centering\arraybackslash} m{1.2cm} >{\centering\arraybackslash} m{1.2cm} | >{\centering\arraybackslash} m{2.5cm} >{\centering\arraybackslash} m{2.5cm}}
		\hline 
		 & & \multicolumn{4}{c}{Identification} & \multicolumn{2}{c}{Test for orthotropy} \\
		 \hline
		 \rule{0pt}{25pt}
	    	& 2D & $E_1$ & $E_2$ & $\nu_{12}$ & $G_{12}$ & $\displaystyle \frac{\vert \mathbb C_{16} \vert}{\mathbb C_{11}}$ & $\displaystyle \frac{\vert \mathbb C_{26} \vert}{\mathbb C_{11}}$  \\
		\hline 
	      & cross	& 416.57 & 371.34 & 0.192 & 158.37  & 0.11 $\%$  & 0.13 $\%$   \\
	 	\hline
	\end{tabular} 
	\caption{Determined material constants (in GPa) of the entire cross-sectional plane assuming orthotropy and the deviation [$\%$] of the remaining elasticity coefficients from perfect orthotropy.} 	
	\label{tab:Deviation Orthotropy}
\end{Table}

Two additional conclusions can be drawn from the planar pre-analyses. 
\begin{itemize}
 \item First, the resultant 3D microstructure is expected to be transverse isotropic due to the identified isotropy in one plane and orthotropy in a perpendicular plane. 
 \item Second, by virtue of isotropic elasticity in the (23) plane the present (12) cross sectional data is representative for the (13) cross sectional plane likewise.  
\end{itemize}
 As a result, microstructure data for three mutually perpendicular planes are available, which is sufficient for the reconstruction of the full 3D microstructure.

\subsection{Identification of representative area elements (RAE)}
 
The full image information of Fig. \ref{fig:Plane-pics-Surf-and-CrossSect} is not adequate for 3D microstructure reconstruction for two reasons. First, since the sizes of the two exemplars are not consistent with each other. Second, and more important, the reconstruction in full resolution would lead to a voxel number well above 200 millions (303$\times$1024$\times$690) which is too expensive for reconstruction and homogenization. Moreover, this sample size is not necessary, since much smaller statistically representative area elements (RAE) can be identified. 
In the present context the selection criterion of an RAE is its statistical representativity in terms of phase fractions and elasticity characteristics as identified above, see Tables~\ref{tab:Homogenized-Elasticity-2d}--\ref{tab:Deviation Orthotropy}. 

Figure \ref{fig:RVE-original-vs-model} displays three selected subwindows of edge length 3\,$\mu$m and their binary images of resolution 270$\times$270 pixels with pixel size 11.1~nm for the (12) cross sections, and of resolution 540 $\times$ 540 pixels with pixel size 5.5~nm for the (23) surface plane. Their placement in the full planar windows is shown by highlighted areas in Fig.~\ref{fig:Plane-pics-Surf-and-CrossSect-Window-Si}.  

As it turns out, the characteristics of the full planar information are accurately captured by the selected RAEs, see Tab.\ref{stiffnes_matrix_RAE}; the deviation of the phase fractions from the full pictures is smaller than 0.05 \%. The deviation from the elasticity coefficients is 0.5\% at maximum if merely the non-vanishing coefficients are considered.

The assumed elasticity laws are captured in good agreement with the full windows, see Tab.~\ref{stiffnes_matrix_RAE}. The assumptions of isotropy and orthotropy, respectively, are underpinned by the excellent agreement of the homogenized coefficients with the idealized ones of the assumed type of material symmetry, see Tabs.~\ref{tab:Deviation RAE Isotropy} and \ref{tab:Deviation RAE Orthotropy}, respectively.

\begin{Figure}[htbp]
	\centering
	 \includegraphics[width=0.49\linewidth]{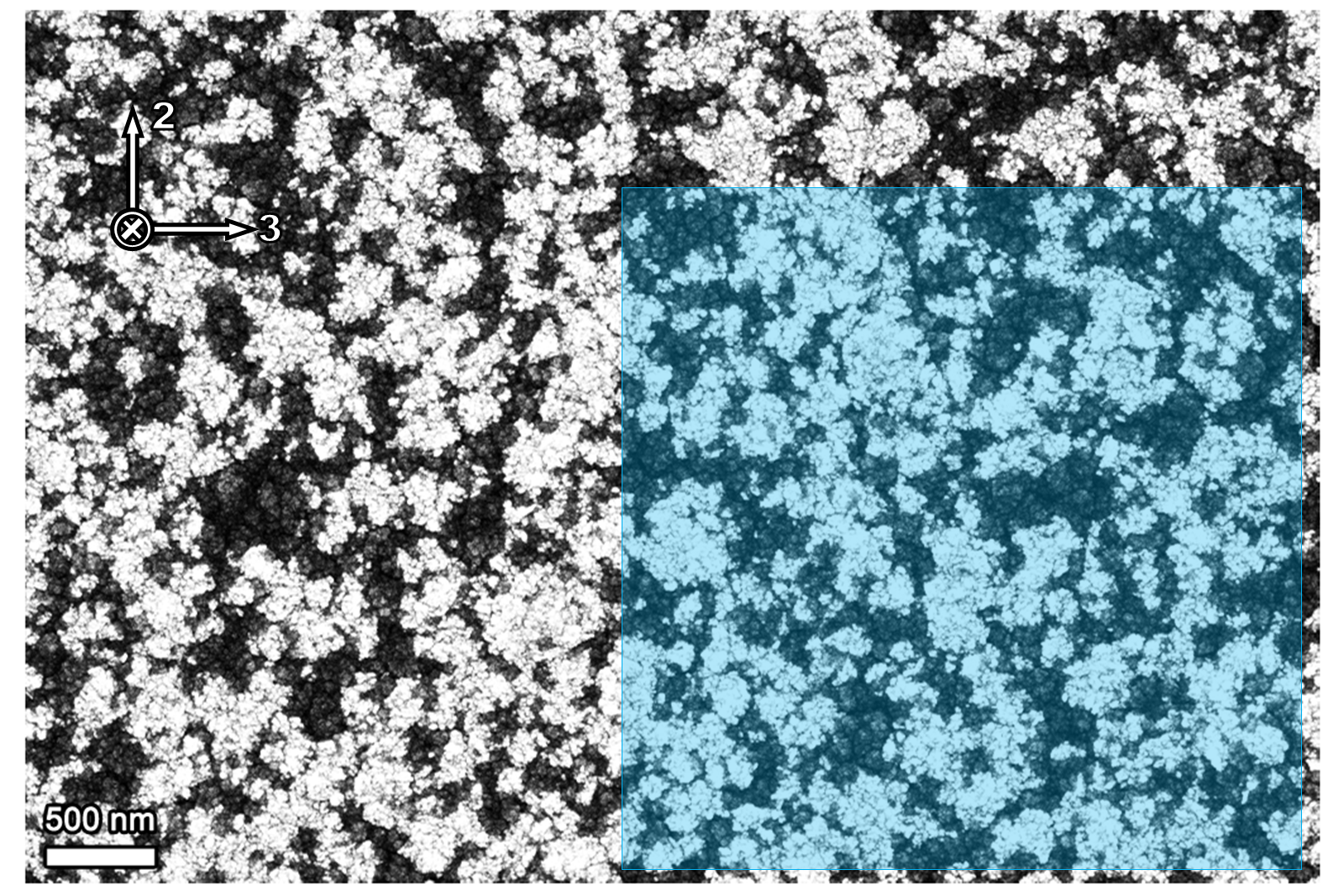}  \hspace*{0mm}
	 \includegraphics[width=0.49\linewidth]{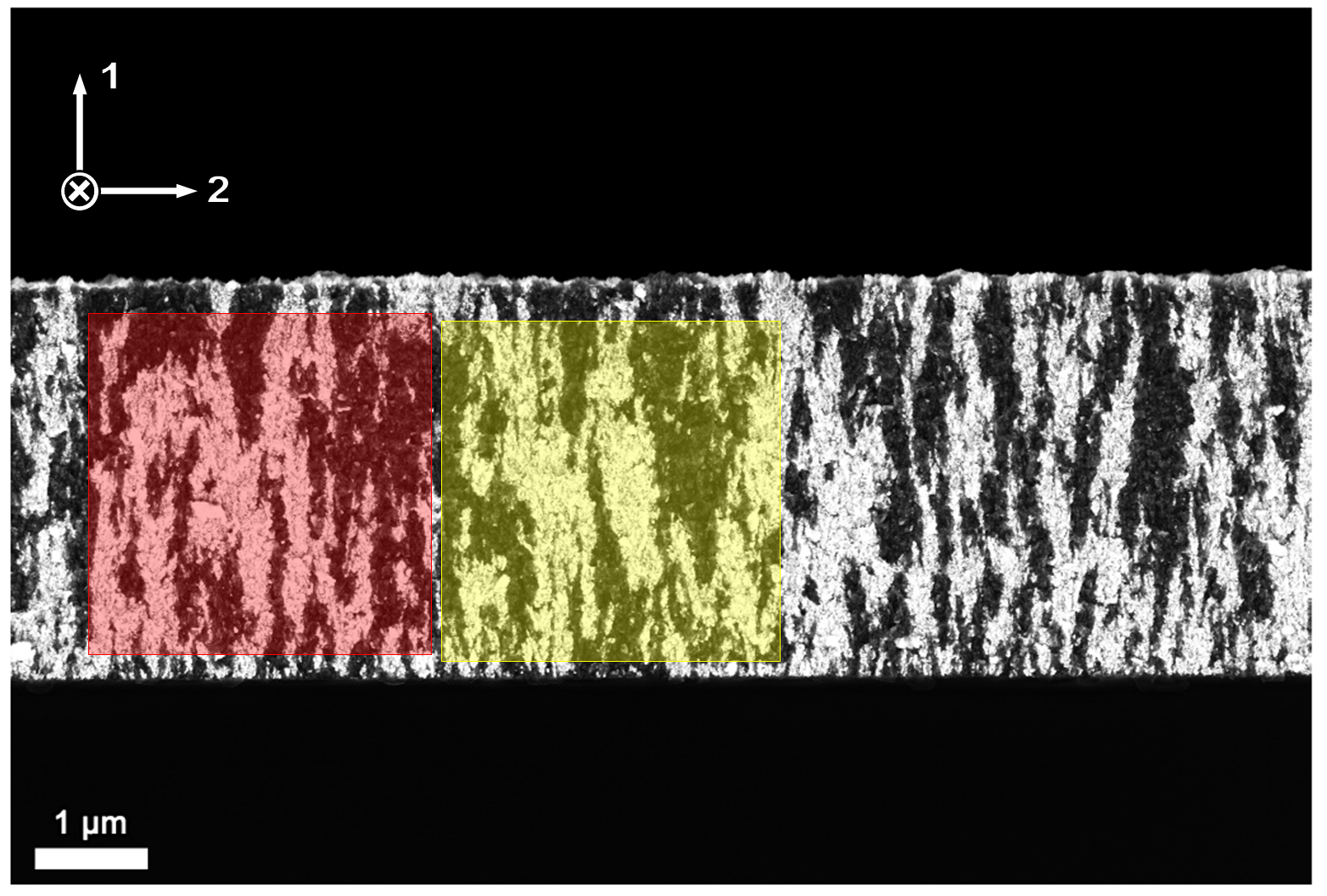}
	\caption{Diamond/$\beta$-SiC composite film: the selected RAEs are highlighted. The coordinate axes define the (23) surface plane  and the (12) cross-sectional plane. \label{fig:Plane-pics-Surf-and-CrossSect-Window-Si}}
\end{Figure}

\begin{Figure}[htbp]
	\centering
	\includegraphics[width=0.1\linewidth]{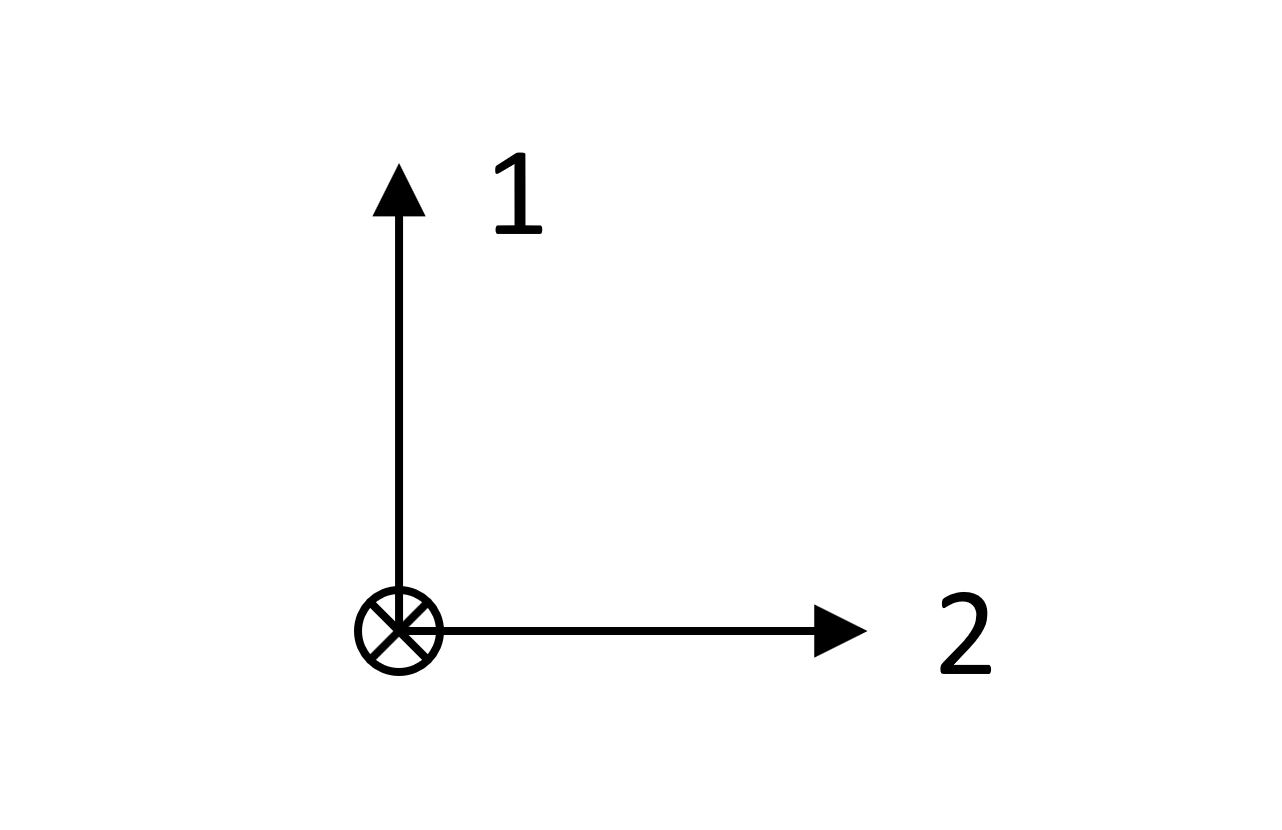}
	\subfigure[]{\includegraphics[height=37mm]{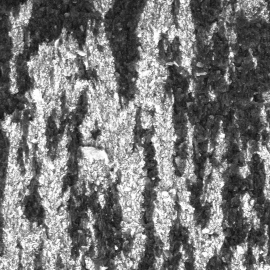}}
	\hspace*{4mm}
	\subfigure[]{\includegraphics[height=37mm]{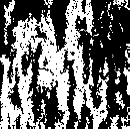}}
	\hspace*{4mm}
	\subfigure[]{\includegraphics[height=37mm]{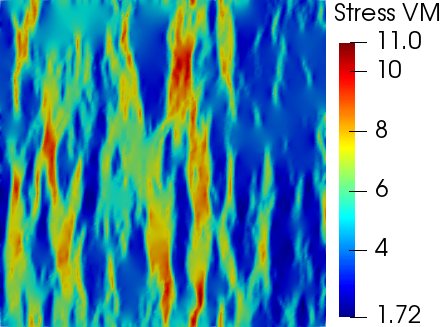}}  
	\\
	\includegraphics[width=0.1\linewidth]{12-kos}
	\subfigure[]{\includegraphics[height=37mm]{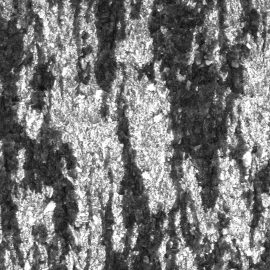}}
	\hspace*{4mm}
	\subfigure[]{\includegraphics[height=37mm]{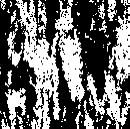}}
	\hspace*{4mm}
	\subfigure[]{\includegraphics[height=37mm]{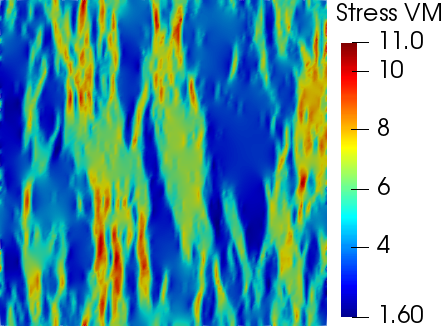}}
	\\
	\hspace*{1.5mm}
	\includegraphics[width=0.1\linewidth]{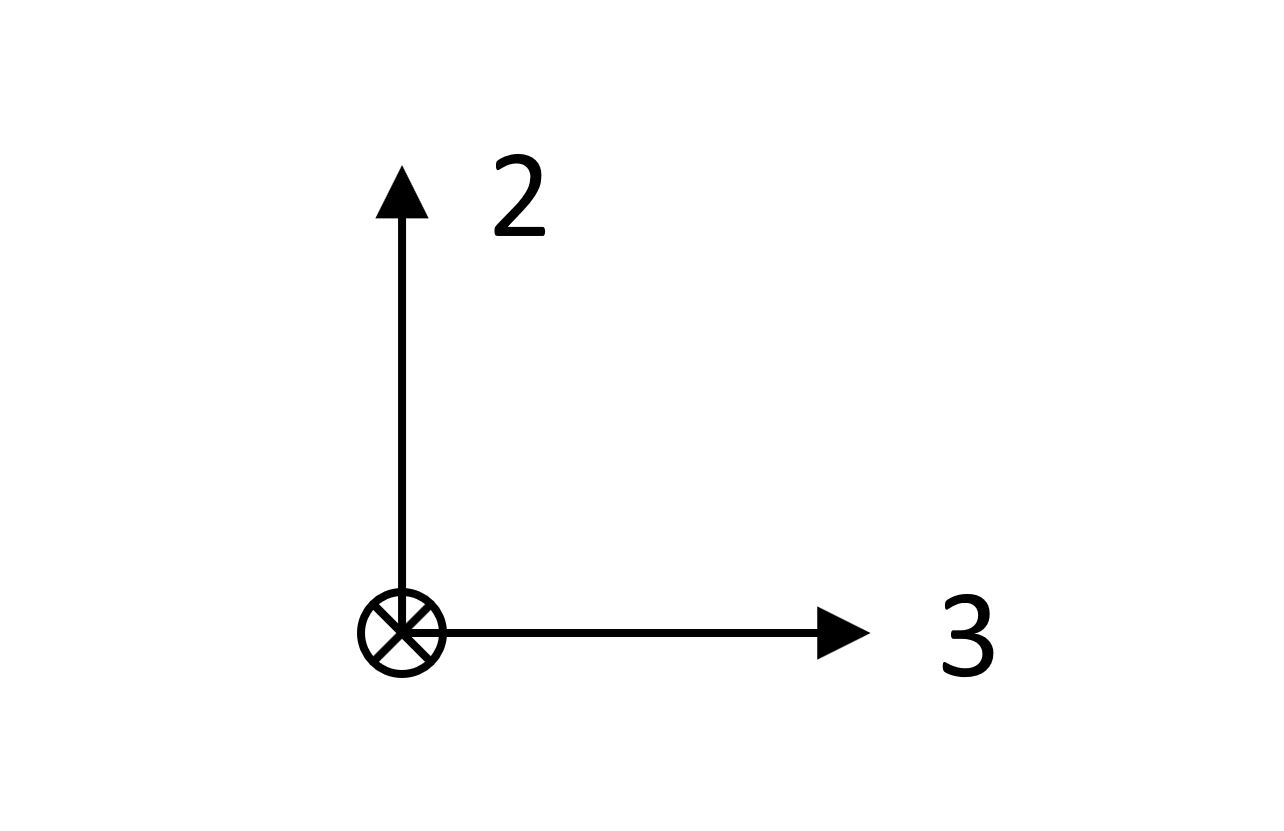}
	\subfigure[]{\includegraphics[height=37mm]{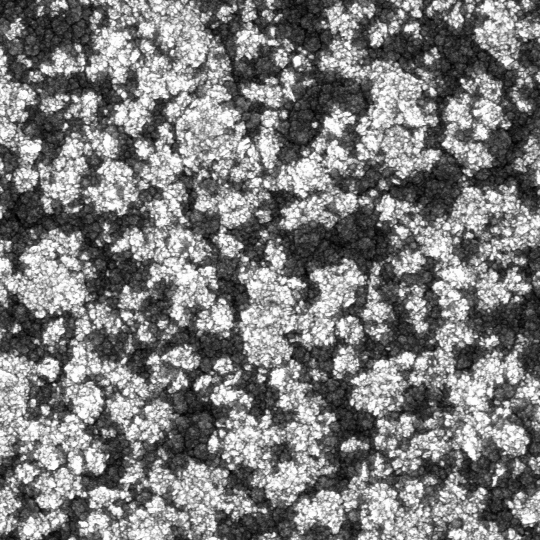}}
	\hspace*{4mm}
	\subfigure[]{\includegraphics[height=37mm]{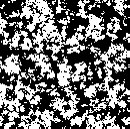}} 
	\hspace*{4mm}
	\subfigure[]{\includegraphics[height=37mm]{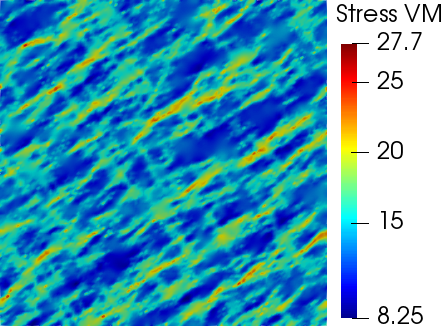}}  
	\caption{Left (a,d,g): SEM micrographs of RAEs with edge length 3 $\mu$m, center(b,e,h): binary representation with 130$\times$130 pixels, right (c,f,i) von-Mises stress for the RAEs in the (c,f) cross sections for compression and in the (i) surface plane for shear.}
	\label{fig:RVE-original-vs-model}
\end{Figure}

\begin{Table}[htbp]
\begin{tabular}{>{\centering\arraybackslash} m{8mm} >{\centering\arraybackslash} m{2cm} >{\centering\arraybackslash} m{2cm} >{\centering\arraybackslash} m{4.5cm}  >{\centering\arraybackslash} m{4.5cm} }
\hline
plane & exemplar & phase fraction deviation & $\mathbb C^\text{RAE}$ [GPa] & $\mathbb C^\text{dev}$ [$\%$] \\
\hline
\\[-4mm]
(23) surf & \includegraphics[width=0.1\textwidth]{surf_3mikrom_130p} & $0.01 \, \%$ & 
$\left[\begin{array}{c r r}
408.6 & 78.1 & -0.6 \\ & 410.0 & -0.8 \\ \scriptstyle \text{sym.} &  & 163.7 
\end{array} \right]$ & $\left[\begin{array}{c r r}
\phantom{-}0.0 & \phantom{-}0.1 & -140.9 \\  & 0.1 & -9.3 \\ \scriptstyle \text{sym.} &  & 0.1 
\end{array} \right]$ \\
(12) cross1 & \includegraphics[width=0.1\textwidth]{cross1_3mikrom_130p} & $0.04 \, \%$ &
$\left[\begin{array}{c r r}
384.7 & 74.1 & 0.0 \\   & 428.7 & -0.3 \\ \scriptstyle \text{sym.} &  & 158.6 
\end{array} \right]$ & $\left[\begin{array}{c r r}
-0.2 & -0.2 & \phantom{-}102.8 \\   & 0.5 & 45.9 \\ \scriptstyle \text{sym.} &  & -0.2 
\end{array} \right]$ \\
(12) cross2 & \includegraphics[width=0.1\textwidth]{cross2_3mikrom_130p} & $0.002 \, \%$ &
$\left[\begin{array}{c r r}
382.9 & 74.1 & -1.4 \\   & 430.6 & -2.4 \\ \scriptstyle \text{sym.} &   & 158.4 
\end{array} \right]$ & $\left[\begin{array}{c r r}
\phantom{-}0.3 & -0.1 & -185.4 \\   & 0.1 & -328.7 \\ \scriptstyle \text{sym.} &   & -0.0 
\end{array} \right]$ \\
\hline
\end{tabular}
\caption{Elastic stiffness matrices $\mathbb C^\text{RAE}$ for various RAEs and their deviation $\mathbb C_\text{ij}^\text{dev}:=(\mathbb C_\text{ij}-\mathbb C_\text{ij}^\text{RAE})/\mathbb C_\text{ij}$ from the homogenized elasticity coefficients of the total images.}
\label{stiffnes_matrix_RAE}
\end{Table}

\begin{Table}[htbp]
	\centering
	\renewcommand{\arraystretch}{1.2}
	\begin{tabular}{ >{\centering\arraybackslash} m{0.01mm} >{\centering\arraybackslash} m{1cm} | >{\centering\arraybackslash} m{1.5cm} >{\centering\arraybackslash} m{1.5cm} | >{\centering\arraybackslash} m{2cm} >{\centering\arraybackslash} m{2cm} >{\centering\arraybackslash} m{1.5cm} >{\centering\arraybackslash} m{1.5cm} }
		\hline 
		  & & \multicolumn{2}{c}{Identification} & \multicolumn{4}{c}{Test for isotropy} \\
		 \hline
		 \rule{0pt}{25pt}
	      & 2D	& $E$ & $\nu$ & $\displaystyle \frac{\mathbb C_{11}- \mathbb C_{22}}{\mathbb C_{11}}$ & $\displaystyle \frac{\vert \mathbb C_{33}-G \vert}{G}$ & $\displaystyle \frac{\vert \mathbb C_{13} \vert}{\mathbb C_{11}}$ & $\displaystyle \frac{\vert \mathbb C_{23} \vert}{\mathbb C_{11}}$    \\
		\hline 
	    & surf	& 395.10 & 0.191 & 0.34 $\%$  & 1.31 $\%$ & 0.14 $\%$  & 0.20 $\%$   \\
	 	\hline
	\end{tabular} 
	\caption{Material parameter identification (in GPa) for the surface RAE under the assumption of isotropy and the deviation [$\%$] of the remaining elasticity coefficients from isotropy.} 	
	\label{tab:Deviation RAE Isotropy}
\end{Table}

\begin{Table}[htbp]
	\centering
	\renewcommand{\arraystretch}{1.2}
	\begin{tabular}{>{\centering\arraybackslash} m{0.01mm} >{\centering\arraybackslash} m{1.3cm} | >{\centering\arraybackslash} m{1.2cm} >{\centering\arraybackslash} m{1.2cm} >{\centering\arraybackslash} m{1.2cm} >{\centering\arraybackslash} m{1.2cm} | >{\centering\arraybackslash} m{2.5cm} >{\centering\arraybackslash} m{2.5cm}}
		\hline 
		  & & \multicolumn{4}{c}{Identification} & \multicolumn{2}{c}{Test for orthotropy} \\
		 \hline
		 \rule{0pt}{25pt}
	      & 2D	& $E_1$ & $E_2$ & $\nu_{12}$ & $G_{12}$ & $\displaystyle \frac{\vert \mathbb C_{13} \vert}{\mathbb C_{11}}$ & $\displaystyle \frac{\vert \mathbb C_{23} \vert}{\mathbb C_{11}}$  \\
		\hline 
	   & cross1	& 414.4 & 371.8 & 0.193 & 158.6  & 0.00 $\%$  & 0.07 $\%$   \\
	   & cross2	& 416.2 & 370.2 & 0.193 & 158.4  & 0.33 $\%$ & 0.57 $\%$   \\
	 	\hline
	\end{tabular} 
	\caption{Identified elasticity constants (in GPa) of the cross-sectional RAEs assuming orthotropy and deviation [$\%$] of the remaining elasticity coefficients from orthotropy.} 	
	\label{tab:Deviation RAE Orthotropy}
\end{Table}

\subsection{Reconstruction of the 3D microstructure}

The procedure of 3D microstructure reconstruction based on the three cross-sectional exemplars in mutually perpendicular planes shall be described. The reconstruction makes use of the 2D exemplars spatial correlations. Methods and software are described in detail in \cite{TurnerKalidindi2016}. 

The RAEs for reconstruction exhibit the edge length of 3 $\mu$m with a consistent resolution of 130 $\times$ 130 pixel (pixel size 23 nm) which results in a final reconstruction volume size of $130^3$ voxels.

Exemplarily, a convergence study for the surface RAE at edge length 3 $\mu$m using various pixel resolutions is carried out, see the appendix, Sec.~\ref{sect:Miscellaneous}; it turns out that the chosen resolution of 130 pixel per edge slightly overestimates the fully converged homogenized elasticity coefficients.

Following a successive hierarchy of a Gaussian pyramid, the exemplars are downsampled two times to half of the resolution in each reconstruction stage. Starting at the coarsest spatial resolution, this result is used as an initial guess for the algorithm at the next level. Alternating between two steps called
the search and the optimization steps until the reconstruction converges (reconstruction stops changing significantly), a statistically representative 3D microstructure is obtained. The goal of the search step is to simply identify the best matching 2D neighborhoods (small square region from the available 2D exemplars) for every spatial bin in the reconstruction. The subsequent optimization step considers all the multiple recommendations and arrives at a single update for the current iteration of the reconstruction \cite{TurnerKalidindi2016}. In doing so, a weighting scheme based on the neighborhood histogram helps to prevent excessive overusing of certain neighborhoods, that drive the algorithm into poor local optima. This constraining of the reconstruction problem with its multiple solutions and local optima to a search space of reasonable solutions can be seen as a sort of regularization.   

The diamond phase fraction of the resultant RVE is 0.43 and thereby well matching the entire surface (0.44) and cross section (0.42) images. The characteristic surface speckles as well as the cross-sectional laminates in the CVD growth direction are present throughout the volume in the corresponding slices.

\begin{Figure}[htbp]  
	\centering
	\subfigure[]{\includegraphics[width=0.32\linewidth]{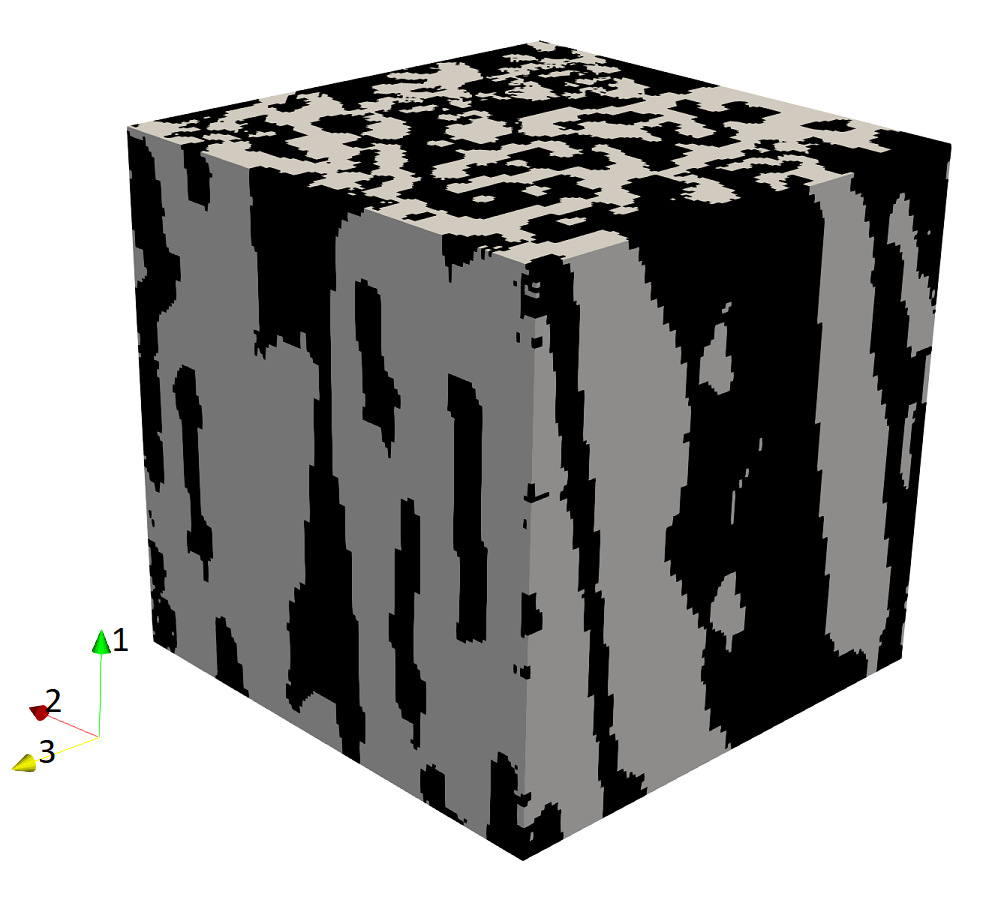}}
	\subfigure[]{\includegraphics[width=0.32\linewidth]{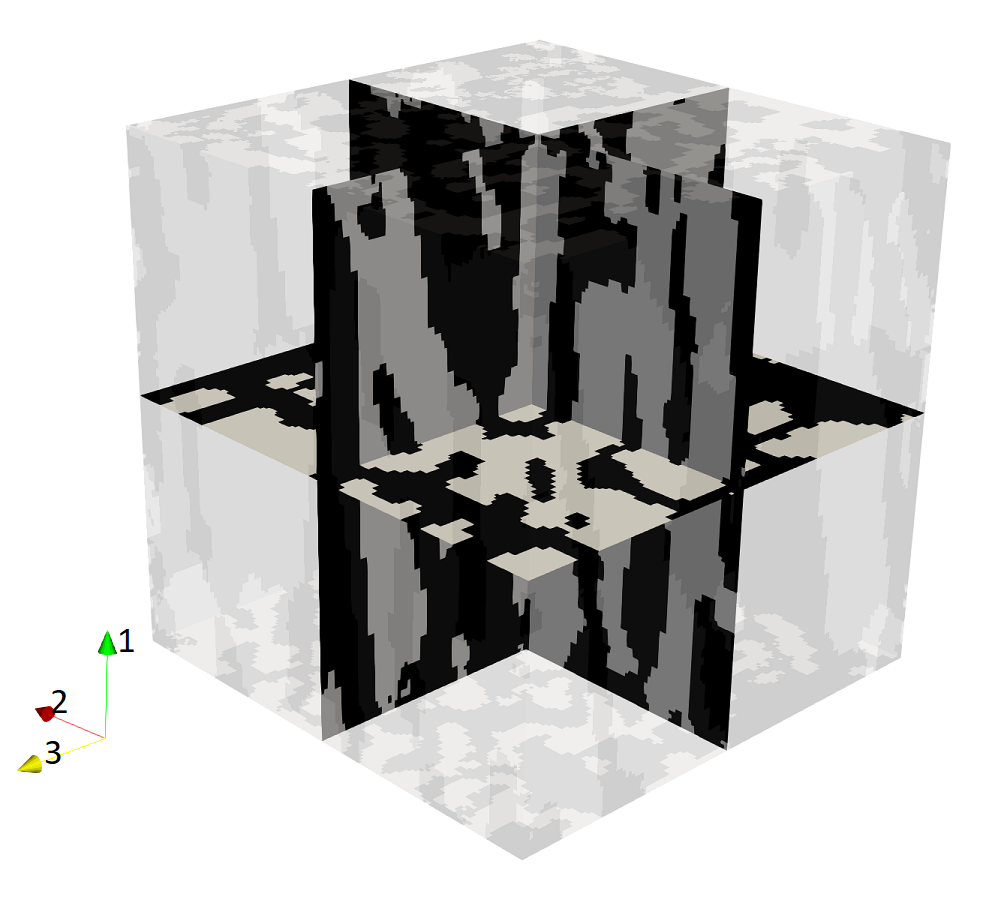}}
	\subfigure[]{\includegraphics[width=0.32\linewidth]{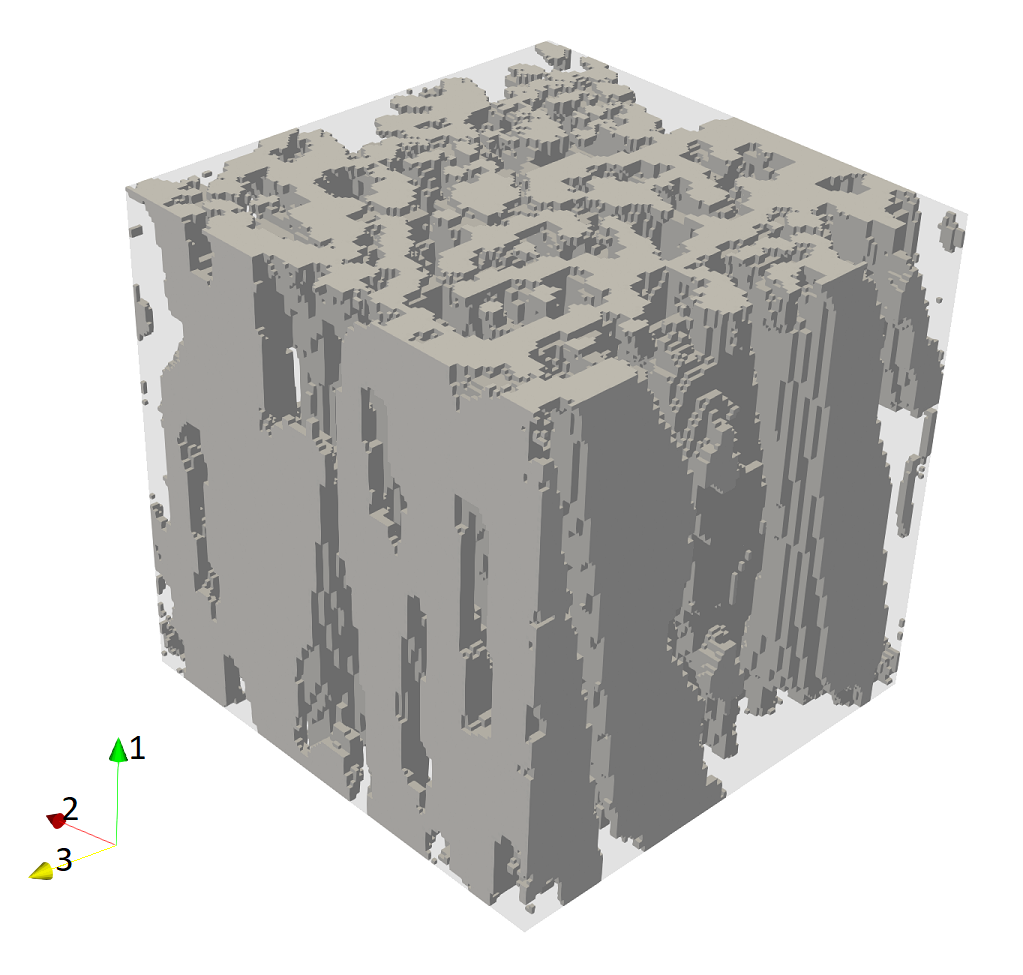}}
	\caption{Reconstructed 3D microstructure: (a) surfaces with diamond (grey) and SiC (black) phases, (b) three perpendicular planes and (c) the diamond phase only. Edge length of the cube is 3 $\mu$m with a total voxel number o 130$^3$.}
	\label{fig:Reconstructed RVE}
\end{Figure}

\subsection{Computation of the 3D homogenized elasticity tensor}

The computation of the effective elastic properties of the two-phase composite is carried out by numerical homogenization, which is based on the Hill-Mandel postulate that the stress power of the homogenized material must be equal to the average stress power in the RVE of the heterogeneous material \cite{Hill1963}, \cite{Hill1972}. The method applied for homogenization is the Finite Element Heterogeneous Multiscale Method (FE-HMM), a two scale-finite element method originally introduced in \cite{E-Engquist-2003}. The method was analyzed for elasticity \cite{Assyr2006} and cast into a finite element formulation in \cite{EidelFischer2018}, \cite{FischerEidel2018}. For overviews of FE-HMM see \cite{E-Engquist-Li-Ren-VandenEijnden2007}, \cite{Assyr-etal2012}.   
 
The simulation is carried out imposing periodic boundary conditions on the reconstructed RVE. The material data for the linear elastic behavior of each phase was given in Tab.~\ref{tab:MatParam}. The finite element discretization of the RVE contains 130 cubic finite elements in each direction of space. Table \ref{tab:Discretizations-ndofs} lists the corresponding number of degrees of freedom (ndof) for the 2D as well as the 3D case.   
  
\begin{Table}[htbp]
\centering
\begin{tabular}{crrcc}
\hline
model     & discretization  & ndof    &      &      \\
\hline
2D        & 130$\times$130             & 34\,322      &   &   \\
3D        & 130$\times$130$\times$130  & 6\,744\,273 &   &   \\
\hline
\end{tabular}
\caption{Discretizations and number of degrees of freedom (ndof) for the RAEs and for the RVE.}
\label{tab:Discretizations-ndofs}
\end{Table}
 
\begin{equation}
\left[\begin{array}{c}
\varepsilon_{11} \\ \varepsilon_{22} \\ \varepsilon_{33} \\ 2\varepsilon_{23} \\ 2\varepsilon_{13} \\ 2\varepsilon_{12}
\end{array}\right]=\left(\begin{array}{c c c c c c}
\frac{1}{E_1} & -\frac{\nu_{12}}{E_1}    & -\frac{\nu_{12}}{E_1}    &                  &                  &                  \\ 
              & \phantom{-}\frac{1}{E_2} & -\frac{\nu_{23}}{E_2}    &                  &                  &                  \\ 
              &                          & \phantom{-}\frac{1}{E_2} &                  &                  &                  \\ 
              &                          &                          & \frac{1}{G_{23}} &                  &                  \\
              &                          &                          &                  & \frac{1}{G_{12}} &                  \\
              &                          &                          &                  &                  & \frac{1}{G_{12}}   
\end{array}\right) \left[\begin{array}{c}
\sigma_{11} \\ \sigma_{22} \\ \sigma_{33} \\ \sigma_{23} \\ \sigma_{13} \\ \sigma_{12}
\end{array}\right]
\label{3D transverse isotropy}
\end{equation}

\begin{Table}[htbp]
	\centering
	\begin{tabular}{>{\centering\arraybackslash} m{0.01mm} >{\centering\arraybackslash} m{1cm} >{\centering\arraybackslash} m{1cm} >{\centering\arraybackslash} m{1.2cm} >{\centering\arraybackslash} m{1.2cm} >{\centering\arraybackslash} m{1cm} >{\centering\arraybackslash} m{1cm} | >{\centering\arraybackslash} m{1.8cm} >{\centering\arraybackslash} m{1.8cm} >{\centering\arraybackslash} m{1.8cm}}
		\hline 
		  & \multicolumn{6}{c}{Identification} & \multicolumn{3}{c}{Test for transverse isotropy} \\
		 \hline
		 \rule{0pt}{25pt}
	     & $E_1$ & $E_2$ & $\nu_{12}$ & $\nu_{23}$ & $G_{23}$ & $G_{12}$ & $\displaystyle \frac{\mathbb C_{12}-\mathbb C_{13}}{\mathbb C_{12}}$ & $\displaystyle \frac{\mathbb C_{22}-\mathbb C_{33}}{\mathbb C_{22}}$ & $\displaystyle \frac{\mathbb C_{66}-\mathbb C_{55}}{\mathbb C_{66}}$   \\
		\hline 
	     & 492.8 & 436.2 & 0.202 & 0.202 & 165.3 & 173.7 & 0.2 $\%$  & 0.6 $\%$ & 0.3 $\%$     \\
	 	\hline
	\end{tabular} 
	\caption{Two-phase composite: Identified material parameters (in GPa) of transverse isotropy for the reconstructed RVE and the deviation [$\%$] of the remaining elasticity coefficients from those of transverse isotropy.} 	
	\label{tab:TransvIsotr3D-Ident-and-Check}
\end{Table}
 
A comparison of coefficients in the homogenized compliance matrix with those of perfect transverse isotropy in the format of \eqref{3D transverse isotropy} identifies the six independent material constants, which are listed in the left of Tab.~\ref{tab:TransvIsotr3D-Ident-and-Check}. The remaining three nonzero coefficients are used to check the validity of the assumed type of elasticity law by means of the criteria $\mathbb{C}_{12}=\mathbb{C}_{13}, \mathbb{C}_{33}=\mathbb{C}_{22}, \mathbb{C}_{55}=\mathbb{C}_{66}$. 
 
The small deviations for all of the three criteria (not more than 0.6\%) in the right of Tab.~\ref{tab:TransvIsotr3D-Ident-and-Check} underpin that the effective elasticity properties are in excellent agreement with the hypothesized transverse isotropy.

\bigskip

{\bf The porous diamond with SiC phase etched out.} \quad In a modified model we consider the case of porous diamond, where the SiC phase is etched out and thereby reproduce in silico what can be realized by an acid agency (e.g. hydrogen fluoride). For the same reconstructed volume as shown in Fig.~\ref{fig:Reconstructed RVE}~(c) we obtain by the same procedure as before the elasticity parameters of Tab.~\ref{tab:TransvIsotr3D-Ident-and-Check-PorousDiamond}. Based on the parameter identfication for $E_1$, $E_2$, $\nu_{12}$, $\nu_{23}$, $G_{23}$, $G_{12}$ the deviation from perfect transverse anisotropy is more pronounced than for the two-phase case. By its structural properties as a cellular material, the porous diamond is clearly much more compliant than the two-phase diamond/$\beta$-SiC composite.
\begin{Table}[htbp]
	\centering
	\begin{tabular}{>{\centering\arraybackslash} m{0.01mm} >{\centering\arraybackslash} m{1cm} >{\centering\arraybackslash} m{1cm} >{\centering\arraybackslash} m{1.2cm} >{\centering\arraybackslash} m{1.2cm} >{\centering\arraybackslash} m{1cm} >{\centering\arraybackslash} m{1cm} | >{\centering\arraybackslash} m{1.8cm} >{\centering\arraybackslash} m{1.8cm} >{\centering\arraybackslash} m{1.8cm}}
		\hline 
		  & \multicolumn{6}{c}{Identification} & \multicolumn{3}{c}{Test for transverse isotropy} \\
		 \hline
		 \rule{0pt}{25pt}
	     & $E_1$ & $E_2$ & $\nu_{12}$ & $\nu_{23}$ & $G_{23}$ & $G_{12}$ & $\displaystyle \frac{\mathbb C_{12}-\mathbb C_{13}}{\mathbb C_{12}}$ & $\displaystyle \frac{\mathbb C_{22}-\mathbb C_{33}}{\mathbb C_{22}}$ & $\displaystyle \frac{\mathbb C_{66}-\mathbb C_{55}}{\mathbb C_{66}}$   \\
		\hline 
	     & 218.7 & 20.4 & 0.400 & 0.129 & 6.02 & 18.93 & -16.1 $\%$  & -30.6 $\%$ & 6.1 $\%$     \\
	 	\hline
	\end{tabular} 
	\caption{Porous diamond: Identified material parameters (in GPa) of transverse isotropy for the porous RVE and the deviation [$\%$] of the remaining elasticity coefficients from those of transverse isotropy.} 	
	\label{tab:TransvIsotr3D-Ident-and-Check-PorousDiamond}
\end{Table}
  
\section{Discussion, Conclusions} 

\subsection{Discussion} 
The main aim of the present work was the 3D virtual reconstruction of an RVE for a diamond/$\beta$-SiC thin film composite and the determination of its effective elastic properties including the type of anisotropy by numerical homogenization.   
 
{\color{black}
For the presented analysis some assumptions are made which refer to the required properties of a volume to qualify for being an  RVE. According to Hill \cite{Hill1963} a representative volume refers ''to a sample that (a) is structurally entirely typical of the whole mixture on average, and (b) contains a sufficient number of inclusions for the apparent overall moduli to be effectively independent of the surface values of traction and displacement, so long as these values are 'macroscopically uniform'.'' It was shown by Sab in \cite{Sab-1992} that for the existence of an RVE the conditions of microstructure ergodicity and statistical homogeneity must be met. In the same reference it was proved that Hill's second requirement can be fulfilled, but only in the infinite-size limit. As a consequence, any homogenized properties exhibit for the finite-size of a volume necessarily a bias due to the employed boundary conditions.
In practice, periodic boundary conditions (PBC) have turned out to be reasonably accurate even for non-periodic matter which is the reason for using PBC in the present work likewise.}
 
A crucial step in the present work was the analysis of microstructures in planar cross sections. The analysis has determined the phase fractions and has identified the types of elastic anisotropies and thereby enabled a reduction to feasible exemplar sizes, the RAEs. Moreover, the obtained result of elastic isotropy in the surface plane was key to complement the existing data in two perpendicular cross sections by a third one. As a consequence a minimal database for the virtual 3D reconstruction procedure was available. 

Notice that the isotropy result in the surface plane and the orthotropy result in a perpendicular plane imply, that (i) orthotropy exists in \emph{all} planes perpendicular to the surface plane, and that (ii) the resultant 3D elasticity is transverse isotropy. While the qualitiative agreement of the homogenized elasticities in 3D directly follows from mechanical reasoning, the figures of the corresponding elasticity parameters must be identified by numerical homogenization. The type of anisotropy is consistent with the CVD fabrication process. In this regard the presented analysis and its results are not only relevant for the particular material system but representative for a broader class of nano-/microcomposite thin films fabricated by CVD or PVD (Physical Vapor Deposition). 

The reconstruction of a 3D RVE from three planar exemplars by means of numerical optimization techniques exhibits some uncertainty, for a thorough discussion we refer to the original paper of \cite{TurnerKalidindi2016} in which the virtual reconstruction method was introduced and validated for a multitude of different microstructures. The advantage of virtual 3D reconstruction based on a few exemplars is that it is ubiquitously available and that it is not as expensive as nanotomography. For the validation of the virtual 3D microstructure an accurate alternative is nanotomography (FIB, SEM) for cross sectional image acquisition, \cite{Holzer-etal-2004}, \cite{Holzer-Cantoni-2012}, and with an application to open, nanoporous structures \cite{Mangipudi-Holzer-Volkert-2016}.

{\color{black}  
The present work expands the field of nanoheterogeneous composite structures which are analyzed in their topology and mechanical properties by a new material system. So far, many significant contributions have considered (functionally graded) carbon-nanofiber/CNT/graphene reinforced polymer and metal composites and combinations therof, \cite{Tjong-2006}, \cite{Ayatollahi-etal-2011}, \cite{Young-etal2012}, \cite{Hernandez-etal-2012}, \cite{Liew-etal-2015}, {\color{black} \cite{Moghadam-etal-2015}}, \cite{HuangCheng2017}, clay/epoxy nanocomposites \cite{Wang-etal-2005}, \cite{Silani-etal-2014}, {\color{black}\cite{Zabihi-etal-2018}}, to name but a few. This contribution adds the analysis of the diamond/$\beta$-SiC composite thin film with characteristics representative for the broad class of CVD- or PVD-based materials.
}
 
\subsection{Conclusions} The simulation results indicate that the nanoheterogeneous diamond/$\beta$-SiC composite fabricated by CVD exhibits the following effective properties  \\[-8mm]
\begin{itemize}  
 \item isotropic elasticity in the surface plane, \\[-7mm]
 \item orthotropic elasticity in \emph{all} planes perpendicular to the surface plane, \\[-7mm]
 \item a resultant transverse isotropic elasticity of the reconstructed 3D microstructure. The corresponding six independent material parameters were identified; a consistency check resulted in minor deviations from those of ideal transverse isotropy with maximal deviation not more than 0.6\%. 
\end{itemize}

The homogenized elasticity tensor opens the door to single-scale finite element analyses of the deformation of diamond/$\beta$-SiC structures at larger length scales, which are cheap and fast. This is a first step towards modeling the rather brittle behavior of this material class up to failure. For that aim, the  experimental identification of damage and failure mechanisms on the submicron length scale is required, its modeling by continuum constitutive laws, and a validation of the latter by two-scale finite element simulations.
 
The methods in this work set the basis to enable future material optimization, where the microstructure of the composite is tailor-made by an appropriate choice of fabrication parameters. By virtue of the simulation methods the optimal parameters (virtual materials testing) can be determined at moderate experimental efforts. 
 
\bigskip

{\bf Declarations of interest:} None. 

{\bf Funding:} This research did not receive any specific grant from funding agencies in the public, commercial, or
not-for-profit sectors.

{\bf Acknowledgments:} BE acknowledges support by the Deutsche Forschungsgemeinschaft (DFG) within the Heisenberg program (grant no. EI 453/2-1). Simulations were performed with computing resources granted by RWTH Aachen University under project ID prep0005. 


\begin{appendix}

\addcontentsline{toc}{section}{Appendix}
\renewcommand{\thesubsection}{\Alph{section}.\arabic{subsection}}
\renewcommand{\theequation}{\Alph{section}.\arabic{equation}}
\renewcommand{\thefigure}{\Alph{section}.\arabic{figure}}
\renewcommand{\thetable}{\Alph{section}.\arabic{table}}
\newcommand {\ssectapp}{
                        \setcounter{equation}{0}
                        \setcounter{figure}{0}
                        \setcounter{table}{0}
		                \subsection
                        }

\setcounter{equation}{0}

\input{appendix_Miscellaneous}

\end{appendix}

%% file: Frontpage-1.tex
\vspace{-3mm}
\begin{center}
  {\bf \large Estimating the Effective Elasticity Properties}\\[2mm]
  {\bf \large of a Diamond/$\beta$-SiC Composite Thin Film} \\[2mm]
  {\bf \large by 3D Reconstruction and Numerical Homogenization} 
\end{center}

\vspace{4mm}
\ce{Bernhard Eidel$^{a,\ast}$, Ajinkya Gote$^{a}$, Marius Ruby$^{a}$, Lorenz Holzer$^{b}$, Lukas Keller$^{b}$, Xin Jiang$^{c}$}   
  
\vspace{4mm}

\ce{\small ${}^a$ Heisenberg-Group, Department Mechanical Engineering} 
\ce{\small University of Siegen, Paul-Bonatz-Str. 9-11, 57068 Siegen, Germany} 
\ce{\small e-mail: bernhard.eidel@uni-siegen.de} 
\vspace{2mm}
\ce{\small ${}^b$ Institute of Computational Physics, School of Engineering} 
\ce{\small Zurich University of Applied Sciences, Wildbachstrasse 21, 8400 Winterthur, Switzerland} 
\ce{\small e-mail: lorenz.holzer@zhaw.ch, lukas.keller@zhaw.ch} 
\vspace{2mm}
\ce{\small ${}^c$ Institute of Materials Engineering, Department Mechanical Engineering} 
\ce{\small University of Siegen, Paul-Bonatz-Str. 9-11, 57068 Siegen, Germany}
\ce{\small e-mail: xin.jiang@uni-siegen.de} 
\bigskip

\bigskip

\begin{center}
{\bf \large Abstract}

\bigskip

{\footnotesize
\begin{minipage}{14.5cm} 
\noindent
The main aim of the present work is to estimate the effective elastic stiffnesses of a two-phase diamond/$\beta$-SiC composite thin film that is fabricated by chemical vapor deposition. The parameters of linear elasticity are determined by numerical homogenization. The database is sparse since for the 3D volume of interest only two micrographs displaying the phase distributions in perpendicular planes are available; micrographs each of a cross-section and the surface of the thin film. A representative volume element (RVE) is reconstructed by an optimization software and by means of identified material symmetries in 2D of the specimen. The elastic homogenization results indicate that the two-phase diamond/$\beta$-SiC composite exhibits the behavior of transverse isotropy, for which the set of six independent material parameters is identified. 
\end{minipage}
}
\end{center}

{\bf Keywords:}
Nanocomposites; Homogenization; Elasticity; Anisotropy; Microstructure reconstruction 

%% file: appendix_Miscellaneous.tex
 
\sect{\color{black} Appendix}
\label{sect:Miscellaneous}

\subsection{Convergence study}
\label{subsec:ConvStudy}

A convergence study of the surface and cross sectional RAEs at edge length 3 $\mu$m using various pixel resolutions reveals that the chosen resolution 
of 130 pixels per edge slightly overestimates the elastic stiffnesses in homogenization, see Fig.~\ref{fig:Convergence-coefficients}. Note that the off-diagonal coefficients do not converge strictly uniform.

\begin{Figure}[htbp]
	\centering  
	\includegraphics[width=0.48\linewidth]{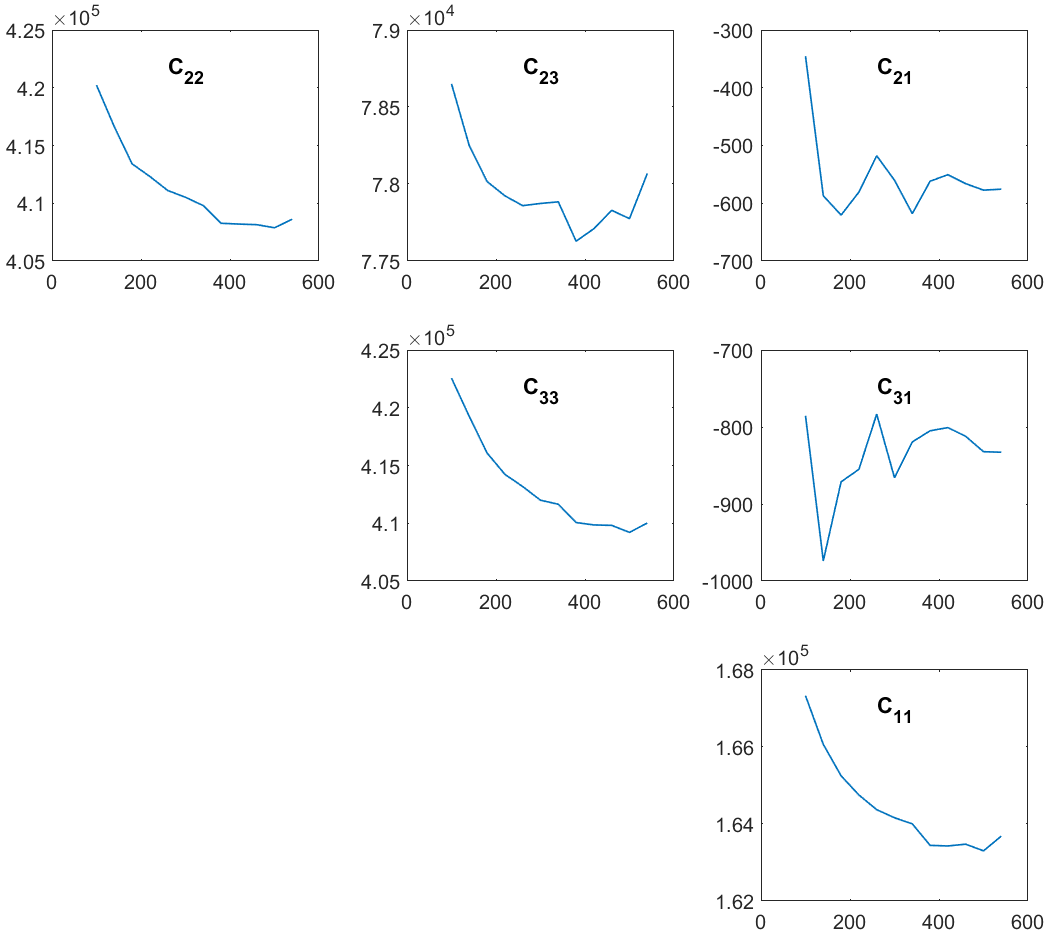} \hspace*{2mm}
	\includegraphics[width=0.48\linewidth]{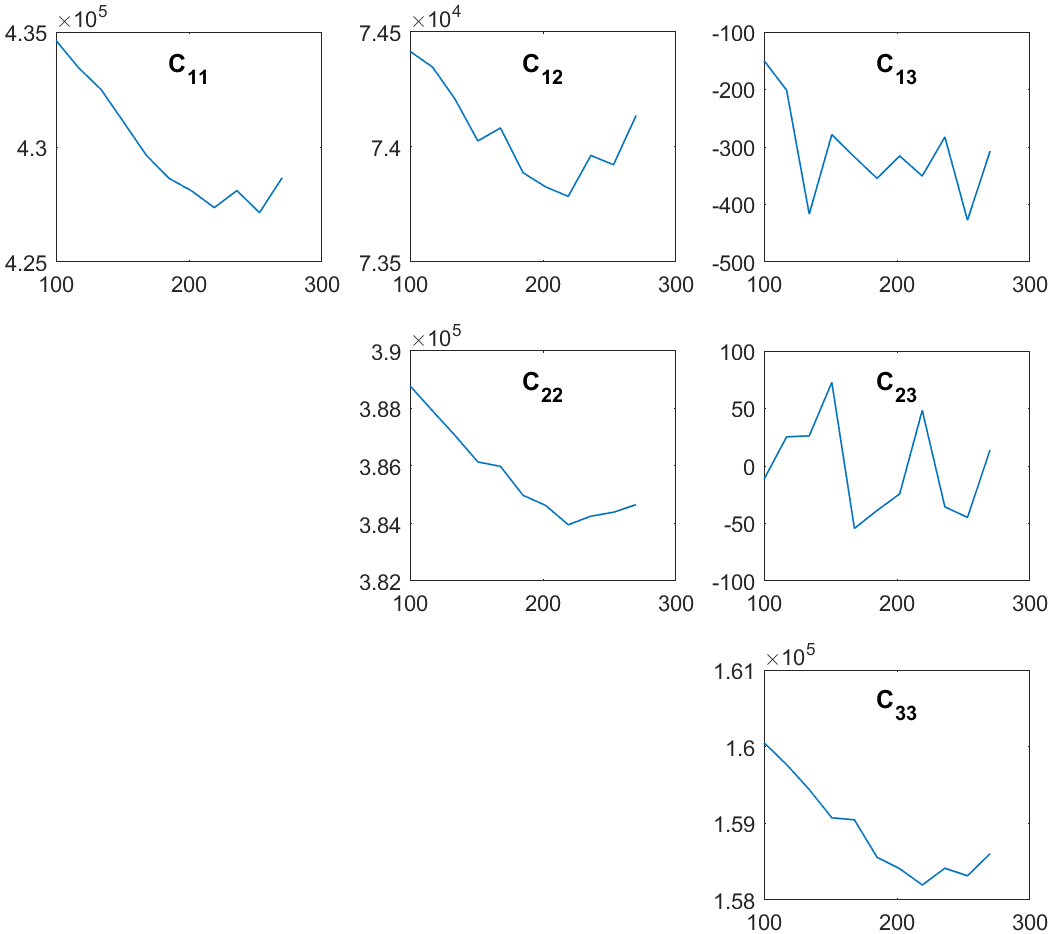}
	\caption{Convergence of homogenized elasticity coefficients as function of the pixel resolution for the exemplars in (left) the surface and 
	(right) the cross section.}
	\label{fig:Convergence-coefficients}
\end{Figure}
 